\documentclass[aps,showpacs,floatfix,prb,twocolumn]{revtex4}
\usepackage{amssymb}
\usepackage{graphicx}
\usepackage{dcolumn}
\usepackage{bm}
\usepackage[]{amsmath}

\begin{document}

\title{Spin-orbit coupling induced Mott transition in Ca$_{2-x}$Sr$_{x}$RuO$_{4}$ (0$\le$ x$\le$ 0.2)}
\author{Guo-Qiang Liu}
\affiliation{Max-Planck-Institut f\"ur Festk\"orperforschung, D-70569 Stuttgart, Germany}
\date{\today }

\begin{abstract}
We propose a new mechanism for the paramagnetic metal-insulator
transition in the layered perovskite Ca$_{2-x}$Sr$_{x}$RuO$_{4}$
(0$\le$x$\le$0.2). The LDA+$U$ approach including spin-orbit
coupling is used to calculate the electronic structures. In
Ca$_{2}$RuO$_{4}$, we show that the spin-orbit effect is strongly
enhanced by the Coulomb repulsion, which leads to an insulating
phase. When Ca is substituted by Sr, the effective spin-orbit
splitting is reduced due to the increasing bandwidth of the
degenerate $d_{xz}$ and $d_{yz}$ orbitals. For $x=0.2$, the
compound is found to be metallic. We show that these results are in good agreement
with the experimental phase diagram.

\end{abstract}
\maketitle

The layered perovskite Ca$_{2-x}$Sr$_{x}$RuO$_{4}$ (CSRO) has been
intensely studied during recent years since this series of
compounds exhibits a variety of interesting physical properties as
a function of the Sr concentration $x$. \cite{Maeno, Cao, Braden,
Nakatsuji, Friedt_1, Friedt_2, Kriener, Pavarini}
Sr$_{2}$RuO$_{4}$ is a p-wave superconductor \cite{Maeno,
Mackenzie} with a K$_{2}$NiF$_{4}$-type structure. The
substitution of Ca for Sr causes the RuO$_{6}$ octahedra to
rotate, and start to tilt at $x=0.5$. \cite{Friedt_1} Following
with the structure distortion, CSRO undergoes a series of phase
transition from a paramagnetic metal (0.5$<$x$<$2) to a magnetic
metal (0.2$<$x$<$0.5), and finally to a Mott insulator
(0$<$x$<$0.2). \cite{Nakatsuji} It is unusual that in the Mott
insulating regime the metal-insulator transition temperature
($T_{\texttt{MI}}$) is higher than the N$\acute{\text{e}}$el
temperature ($T_{N}$) of the antiferromagnetic (AFM) phase, which
shows that a paramagnetic (PM) insulating phase exists between
these transition temperatures. \cite{Nakatsuji_2, Alexander,
Friedt_1} For pure Ca$_{2}$RuO$_{4}$, the PM insulating regime
extends  from $T_{N}$= 110 K to $T_{\texttt{MI}}$= 357 K.
\cite{Cao, Nakatsuji_2, Braden} This property makes
Ca$_{2-x}$Sr$_{x}$RuO$_{4}$ (0$<$x$<$0.2) different from other AFM
Mott insulators.

Recently, Qi $et$ $al$. \cite{Qi} found that the substitution of
the lighter Cr for the heavier Ru strongly depresses
$T_{\texttt{MI}}$ in Ca$_{2}$Ru$_{1-y}$Cr$_{y}$O$_{4}$
(0$<$y$<$0.13), which implies a possible influence of the
relativistic spin-orbit (SO) coupling on the Mott transition as
pointed out by the authors. It is well known that SO coupling
plays an important role in $5d$ transition-metal oxides. For
example, Kim $et$ $al$. \cite{Kim} found that Sr$_{2}$IrO$_{4}$ is
a $J_{eff}=1/2$ Mott insulator, and they showed that the unusual
insulating state can be explained by the combined effect of the SO
coupling and Coulomb interaction. In the $4d$ oxides, the
importance of SO coupling is under debate. Mizokawa $et$
$al$.,\cite{Mizokawa} observed strong SO coupling in
Ca$_{2}$RuO$_{4}$ from their photoemission experiment. Based this
finding, they argued that the strong SO coupling in
Ca$_{2}$RuO$_{4}$ would cause a complex electronic configuration.
Theoretical studies revealed strong SO effects in
Sr$_{2}$RuO$_{4}$ and Sr$_{2}$RhO$_{4}$,\cite{Haver, Liu} which
seemingly support the photoemission experiment. However, Fang $et$
$al$.\cite{Fang1, Fang2} reported an LDA+$U$ study of
Ca$_{2}$RuO$_{4}$. They found the AFM state has a rather simple
configuration $xy^{\,\uparrow \downarrow }\,xz^{\,\uparrow
}\,yz^{\,\uparrow }$ without much influence of the SO coupling.
These seemingly inconsistent viewpoints raise a question: what
role does the SO coupling play in CSRO?

In this paper we present electronic structure calculations for
Ca$_{2-x}$Sr$_{x}$RuO$_{4}$ using the LDA+$U$ method including the
SO coupling. We show the combination of the SO coupling and
Coulomb repulsion opens a band gap in PM Ca$_{2}$RuO$_{4}$. The
appearance of the Mott insulating phase is strongly dependent on
the tilting of the RuO$_{6}$ octahedra, which naturally explains
the PM Mott transition in the experimental phase diagram. On the
other hand, we find SO has much less influence on the AFM order.
We show that these phenomena can be explained by a simple
formalism.

All the calculations in this work were performed with the
full-potential linear augmented plane wave (FLAPW) within the
local-density approximation (LDA), as implemented in package
WIEN2K. \cite{wien} Two experimental structure \cite{Friedt_1}
were considered in this work. For Ca$_{2}$RuO$_{4}$, we used the
structure at 180 K, with the space group Pbca, lattice constant
a=5.394, b=5.600, and c=11.765 \AA. \cite{Friedt_1} For
Ca$_{1.8}$Sr$_{0.2}$RuO$_{4}$, we used the experimental structure
at 10 K, but the substitution of Sr for Ca is only taken into
account via the structural changes. Ca$_{1.8}$Sr$_{0.2}$RuO$_{4}$
also has the space group Pbca but with lattice constant a=5.330,
b=5.319, and c=12.409 \AA. \cite{Friedt_1} For the AFM state, we
considered the 'A-centered' mode. \cite{Braden} The LDA+$U$
calculations were performed with $U=3.0$ eV, which is similar to
the value used by Fang $et$ $al$. \cite{Fang1, Fang2} We will show
that this U value can reproduce the measured band gap in
Ca$_{2}$RuO$_{4}$.

\begin{figure}[tbp]
{\scalebox{0.68}[0.68]{\includegraphics{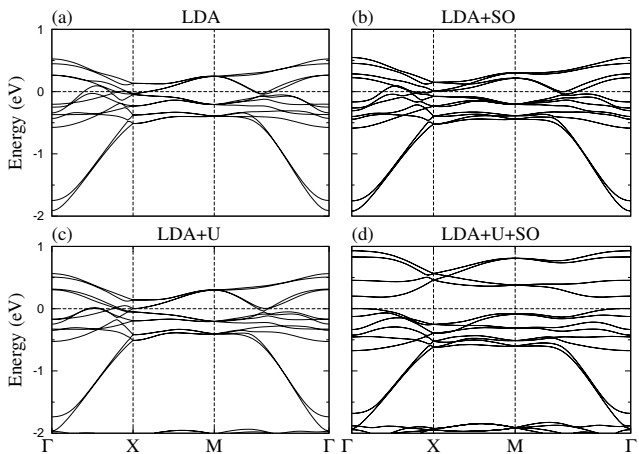}}}
\caption{Theoretical band structures for paramagnetic
Ca$_{2}$RuO$_{4}$ using different approximations, (a) LDA, (b)
LDA+SO, (c) LDA+$U$, and (d) LDA+$U$+SO. The LDA+$U$ and
LDA+$U$+SO band structures are calculated with $U$= 3.0 eV.}
\label{band}
\end{figure}

In Fig. \ref{band}, we present our theoretical band structures for
paramagnetic Ca$_{2}$RuO$_{4}$ using different approximations. The
LDA band structure is well known \cite{Woods, Fang3}: the bands
crossing the Fermi level are from Ru $t_{2g}$ orbitals, containing
four $d$ electrons. Our LDA band structure shown in Fig. 1a is
consistent with the previous study. \cite{Woods} The inclusion of
the SO coupling (Fig. 1b) only shows some slight changes on the
band structure. This is not surprising since the SO coupling
constant $\zeta$ in Ca$_{2}$RuO$_{4}$ is presumably similar to the
one in Sr$_{2}$RuO$_{4}$, where it is only about 93 meV.
\cite{Haver} The inclusion of Coulomb interaction (Fig. 1c) also
shows little influence on the band structure since U does not
break the orbital symmetry in the paramagnetic state.
Surprisingly, the combined interaction of the SO coupling and
Coulomb repulsion gives a very different band structure compared
to the LDA, LDA+SO or LDA+$U$ results.  The LDA+$U$+SO band
structure shows an insulating phase with a gap about 0.2 eV wide.
The band gap obtained from the chosen U is in good agreement with
the experimental data. \cite{Cao, Puchkov}

Similar combined effect of the SO coupling and U has been found in
Sr$_{2}$RhO$_{4}$, \cite{Liu} where it was termed
$Coulomb$-$enhanced$ $spin$-$orbit$ $splitting$. Sr$_{2}$RhO$_{4}$
has a similar crystal structure to Ca$_{2-x}$Sr$_{x}$RuO$_{4}$,
and it can be regarded as a two-band ($xz$ and $yz$) system since
the $xy$ band is below the Fermi level due to the RhO$_{6}$
rotation. \cite{Bkim, Ko} The simpler problem of Sr$_{2}$RhO$_{4}$
can help us to understand the LDA+$U$+SO band structure of
Ca$_{2}$RuO$_{4}$. In Sr$_{2}$RhO$_{4}$, the SO coupling splits
the degenerate $xz$ and $yz$ bands to the higher $\chi_{\pm3/2}$
bands, and lower $\chi_{\pm1/2}$ bands, where
\begin{eqnarray*}
\chi_{3/2}&=&\left( xz+iyz\right) \uparrow, \quad\chi _{-3/2}=\left( xz-iyz\right) \downarrow \nonumber\\
\chi_{1/2}&=&\left( xz+iyz\right) \downarrow, \quad\chi _{-1/2}=\left( xz-iyz\right) \uparrow.\nonumber\\
\end{eqnarray*}
This splitting happens around the Fermi level, and therefore the
occupancies of the $\chi_{\pm3/2}$ and $\chi_{\pm1/2}$ states are
changed: $(n_{1/2}+n_{-1/2})-(n_{3/2}+n_{-3/2})=p>0$, where
$n_{1/2}=n_{-1/2}$ and $n_{3/2}=n_{-3/2}$. When the Coulomb
interaction is taken into account, the SO splitting is enhanced
due to the different occupancies of the $\chi_{\pm3/2}$ and
$\chi_{\pm1/2}$ states. The interplay the SO coupling and Coulomb
interaction can be represented by an effective SO constant
\cite{Liu}
\begin{equation}
\zeta_{eff}=\zeta +\frac{1}{2}(U-J)p,
\end{equation}
where $J$ is the Hund's coupling and $p$ is determined
selfconsistently.

\begin{figure}[tbp]
{\scalebox{0.65}[0.65]{\includegraphics{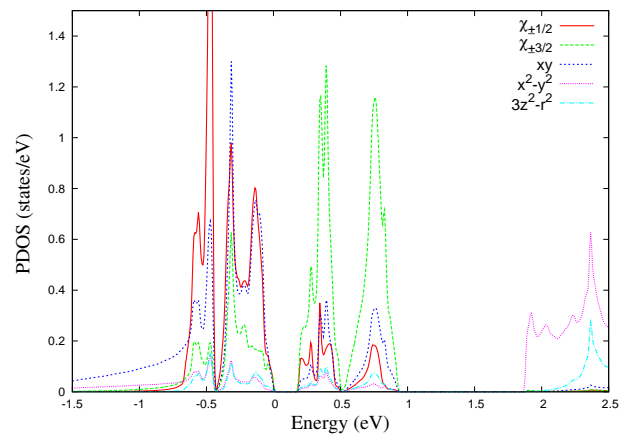}}}
\caption{Ru-$d$ PDOS for paramagnetic Ca$_{2}$RuO$_{4}$ calculated by LDA+$U$+SO.} \label{DOS}
\end{figure}

The problem of Ca$_{2}$RuO$_{4}$ is more complicated than
Sr$_{2}$RhO$_{4}$ since the $xy$ orbital is also involved. Fig.
\ref{DOS} presents the partial density of states (PDOS) for the
Ru-$d$ orbitals calculated by LDA+$U$+SO. Here we present the PDOS
for the $\chi_{\pm3/2}$ and $\chi_{\pm1/2}$ orbitals instead of
$xz$ and $yz$. The PDOS shows that the unoccupied $t_{2g}$ bands
(0.2-0.9 eV) are dominated by the $\chi_{\pm3/2}$ states while the
$\chi_{\pm1/2}$ states are nearly fully occupied. The well
separated $\chi_{\pm3/2}$ and $\chi_{\pm1/2}$ states indicate a
large effective spin-orbit splitting in Ca$_{2}$RuO$_{4}$.
Therefore, A simple explanation for the PM Mott transition is that
the Coulomb-enhanced spin-orbit splitting opens a gap between the
$\chi_{\pm3/2}$ and $\chi_{\pm1/2}$ bands, leading to an
insulating phase with two holes residing on the $\chi_{\pm3/2}$
orbitals. In this explanation, the $xy$ state is assumed to be
fully occupied. However, in the experimental structure, the $xy$,
$xz$ and $yz$ orbitals hybridize with each other due to the
structural distortion. As may be seen, the weight of the $xy$
state in the unoccupied $t_{2g}$ bands is not small as shown in
Fig. \ref{DOS}. The relative hole population shown in Fig.
\ref{DOS} is
$xy$:$\chi_{\pm\frac{1}{2}}$:$\chi_{\pm\frac{3}{2}}$=19:13:68,
while this ratio is 21:39.5:39.5 within LDA approximation. This
shows that the inclusion of SO and U hardly changes the occupancy
of the $xy$ orbital.  We may conclude that the band gap is mainly
due to the splitting of $\chi_{\pm3/2}$ and $\chi_{\pm1/2}$
orbitals although the $xy$ orbital is also involved in the Mott
transition.

Experimental research has found that the Mott transition in CSRO
is accompanied by an structural phase transition from the high
temperature $L$-$Pbca$ phase to the low temperature $S$-$Pbca$
phase, \cite{Friedt_1,Alexander} where L (S) indicates a long
(short) $c$-axis. The phase transition temperature $T_{S}$ is a
function of Sr concentration $x$, which decreases from 357 K at
$x=0$ to 0 K at $x\thicksim0.2$ \cite{Friedt_1,Alexander}. For
$x\ge0.2$, CSRO is metallic and only has the $L$-$Pbca$ phase. As
indicated by Friedt $et$ $al$., \cite{Friedt_1}, the structural
transition from the $L$-$Pbca$ to $S$-$Pbca$ phase is
characterized by an increase in the tilting angle of RuO$_{6}$
octahedra. We will show that the tilting angle of RuO$_{6}$ plays
an important role in the Mott transition. To illuminate the
relation between the Mott transition and the structural phase
transition, we apply the LDA+$U$+SO calculation to
Ca$_{1.8}$Sr$_{0.2}$RuO$_{4}$.

\begin{figure}[tbp]
{\scalebox{0.65}[0.70]{\includegraphics{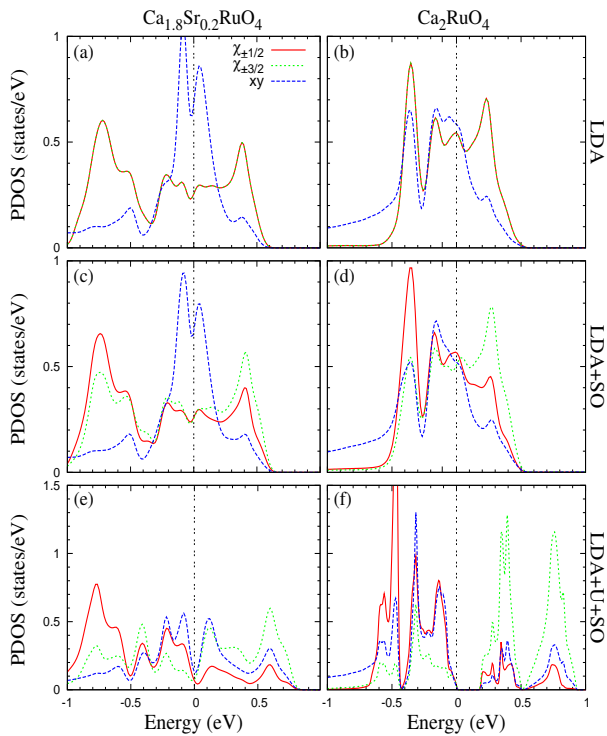}}}
\caption{Ru-$t_{2g}$ PDOS for paramagnetic state using
different approximations. The left panels are for
Ca$_{1.8}$Sr$_{0.2}$RuO$_{4}$, and right panels for
Ca$_{2}$RuO$_{4}$.} \label{tilting}
\end{figure}

Fig. 3 presents our calculated PDOS for $x=0.2$ ($L$-$Pbca$) and
$x=0$ ($S$-$Pbca$). Fig. 3a and 3b show the LDA PDOS for $x=0.2$
and $x=0$. The $xz$/$yz$ bandwidth is about 1.6 eV in
Ca$_{1.8}$Sr$_{0.2}$RuO$_{4}$, and it is reduced to 1.1 eV in
Ca$_{2}$RuO$_{4}$. The narrower $xz$/$yz$ band in
Ca$_{2}$RuO$_{4}$ is due to its larger tilting angle. The tilting
angle is about 12$^{\circ}$ in Ca$_{2}$RuO$_{4}$, and it is
6$^{\circ}$ in Ca$_{1.8}$Sr$_{0.2}$RuO$_{4}$. \cite{Friedt_1} The
tilting of the in-plane Ru-O can significantly reduces the
interaction between Ru-$d_{xz/yz}$ and O-$p_{z}$. Consequently,
the $xz$/$yz$ bandwidth decreases from $x=0.2$ to $x=0$, while the
$xy$ bandwidth is less influenced. With the narrower $xz$/$yz$
band, Ca$_{2}$RuO$_{4}$ shows much higher $xz$/$yz$ PDOS around
the Fermi level than Ca$_{1.8}$Sr$_{0.2}$RuO$_{4}$. Fig. 3c and 3d
show the LDA+SO PDOS for $x=0.2$ and $x=0$. As may be seen, the
occupancy difference between the $\chi_{\pm3/2}$ and
$\chi_{\pm1/2}$ states is larger in Ca$_{2}$RuO$_{4}$ than in
Ca$_{1.8}$Sr$_{0.2}$RuO$_{4}$. This is understandable if we
consider the higher PDOS in Ca$_{2}$RuO$_{4}$. Eq. (1) shows the
effective SO splitting is proportional to the occupancy difference
$p$. Then the lager occupancy difference $p$ in Ca$_{2}$RuO$_{4}$
will cause larger SO splitting when Coulomb interaction is taken
into account. This is confirmed by the LDA+$U$+SO PDOS shown in
Fig. 3e and 3f. Using Eq. (1), we get $\zeta_{eff}$=0.9 eV for
Ca$_{2}$RuO$_{4}$, and 0.6 eV for Ca$_{1.8}$Sr$_{0.2}$RuO$_{4}$.
The larger SO splitting in Ca$_{2}$RuO$_{4}$ leads an insulating
phase, while Ca$_{1.8}$Sr$_{0.2}$RuO$_{4}$ remains metallic.
Therefore, we have shown that the PM Mott transition in CSRO can
be explained by the interplay of SO coupling, Coulomb interaction
and structural distortion.

CSRO is a single-layer system, where the $t_{2g}$ bands are split
into the singly degenerate $xy$ band and doubly degenerate
$xz$/$yz$ bands. The bare SO coupling mainly influences the
degenerate bands since Ru has a moderate SO constant. Therefore,
the SO induced Mott transition is strongly orbital dependent. Fig.
3 indicates that the Mott transition in PM CSRO is driven by the
narrowing of the $xz$ and $yz$ bands, while the $xy$ orbital plays
a less important role. This strong orbital-dependence could also
be seen from Eq. (1). The bare SO parameter $\zeta$ is constant
for each orbital, but the Coulomb enhanced SO parameter
$\zeta_{eff}$ is a function of orbital occupancies. This suggests
that the strong orbital-dependence is an inherent feature of the
SO induced Mott transition.

\begin{figure}[tbp]
{\scalebox{0.65}[0.65]{\includegraphics{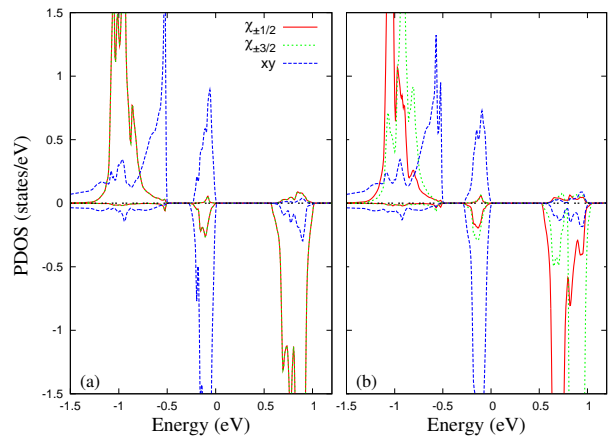}}}
\caption{Ru-$t_{2g}$ PDOS for AFM Ca$_{2}$RuO$_{4}$ using
different approximations, (a) LDA+$U$, and (b) LDA+$U$+SO.}
\label{AFM}
\end{figure}

Our calculation has shown that the PM insulating phase of CSRO
originates in the strong effective SO splitting. This picture
supports the photoemission measurement by Mizokawa
$et$ $al$. \cite{Mizokawa} The suppression of $T_{\texttt{MI}}$ in
Ca$_{2}$Ru$_{1-y}$Cr$_{y}$O$_{4}$ \cite{Qi} can also be understood
within this picture. Cr has a much smaller atomic SO constant than
Ru due to its smaller mass. And therefore the substitution of Cr for Ru
will reduce the SO splitting, leading to the observed decrease of $T_{\texttt{MI}}$.

As mentioned above, Fang $et$ $al$. \cite{Fang1, Fang2} found that
SO coupling has no much influence on the electronic configuration.
They however pointed out that the photoemission measurement was
done above the N$\acute{\text{e}}$el temperature, while they
applied the LDA+$U$ method to the low temperature AFM state. To
clarify if the SO coupling is less important in AFM state, we
apply the LDA+$U$ and LDA+$U$+SO calculation to AFM
Ca$_{2}$RuO$_{4}$. Our LDA+$U$ calculation gives a magnetic moment
of $m_{Ru}$=1.25 $\mu_{B}$, which is consistent with Fang $et$
$al$.'s calculation \cite{Fang1}; while SO reduces the moment to
1.21 $\mu_{B}$, showing a weak SO effect. The AFM PDOS for
Ca$_{2}$RuO$_{4}$ are presented in Fig. 4. In contrast to the PM
state, Fig. 4 shows that there is no Coulomb-enhanced SO splitting
in the AFM state. The relative weak SO splitting in the AFM state
can be explained by Eq. (1). The LDA+$U$ calculation produces an
insulating phase for AFM Ca$_{2}$RuO$_{4}$ as shown in Fig. 4a.
Since there is no density of states around the Fermi level, SO
coupling can not change the orbital occupancies, which gives
$p=0$. Then we get $\zeta_{eff}=\zeta$, showing no enhanced SO
splitting. It is noticeable that our calculations give an AFM
ground state for Ca$_{2}$RuO$_{4}$, which is consistent with the
experimental phase diagram. \cite{Friedt_1}

Comparing the PM and AFM insulating phases in CSRO, we may find
that the two kinds of Mott transition are similar. They both have
an interaction to break the orbital symmetry. The interaction is
SO coupling in the PM state and spin polarization in the AFM
state. The breaking of orbital symmetry lifts the degenerate bands
and changes the orbital occupancies. When the Coulomb interaction
is taken into account, the orbital splitting, which is SO
splitting in the PM state or exchange splitting in the AFM state,
is enhanced. If the enhanced splitting is large enough, it will
lead to an insulating phase. The PM-AFM transition at $T_{N}$ can
be regarded as the competition of the Coulomb-enhanced SO
splitting and the Coulomb-enhanced exchange splitting. In the AFM
state, the SO enhancement is quenched by the large exchange
splitting, which causes the very different electronic
configuration from the PM state.

In summary, we have applied LDA+$U$+SO calculations to CSRO. We
find the Coulomb enhanced SO splitting produces an insulating
phase in PM Ca$_{2}$RuO$_{4}$. This finding is consistent with the
photoemission experiment, and also explains the recent experiment
on Ca$_{2}$Ru$_{1-y}$Cr$_{y}$O$_{4}$. We show that the SO induced
Mott transition in CSRO is driven by the change of the $xz$/$yz$
bandwidth. For $x=0.2$, the compound is found to be metallic. On
the other hand, we find that SO coupling has much less influence
on the AFM state, which is in agreement with the previous LDA+$U$
study. The above picture shows that SO coupling plays a very
subtle role in the correlated systems. The interplay of SO
coupling, electron correlation and crystal structure distortion
would cause very rich physical phenomena.

The author gratefully acknowledges Ove Jepsen for helpful
discussions and useful comments.

\end{document}